\documentclass[a4paper,12pt]{article}
\usepackage{amsmath}
\usepackage{pstricks}
\usepackage{pst-node}
\usepackage[ansinew]{inputenc}
\usepackage{amssymb,amsmath}
\usepackage{amsfonts}
\usepackage{epsfig}

\def\p{\partial}
\def\px{\partial_x}
\def\py{\partial_y}
\def\a{\alpha}
\def\b{\beta}
\def\g{\gamma}

\def\o{\omega}

\font\Sets=msbm10

\def\Integer {\hbox{\Sets Z}}    \def\Real {\hbox{\Sets R}}

\def\Complex {\hbox{\Sets C}}

\def\be{\begin{equation}}       \def\ba{\begin{array}}

\def\ee{\end{equation}}         \def\ea{\end{array}}

\def\bea {\begin{eqnarray}}      \def\eea {\end{eqnarray}}

\def\bean{\begin{eqnarray*}}    \def\eean{\end{eqnarray*}}

\def\const {\mathop{\rm const}\nolimits}

\def\im    {\mathop{\rm Im}   \nolimits}

\def\ker  {\mathop{\rm Ker} \nolimits}

\def\qed   {\vrule height0.6em width0.3em depth0pt}

\def\<{\langle} \def\({\left(}  \def\>{\rangle} \def\){\right)}

\newtheorem{exi}{Example}

\author{Alexey Shabat$^{1}$, Elena Kartashova$^2$\\\\
$^{1}$ Lecce University, Lecce, Italy\\
$^2$ RISC, J.Kepler University, Linz, Austria\\
\\\\  e-mail: lena@risc.uni-linz.ac.at
}

\title{Computable Integrability.\\
 Chapter 1: General notions and ideas\footnote{This is a preliminary
version of the  Chapter 1 of a planned book "Computable
Integrability."}}

\begin{document}
%generates titel
\date{}
\maketitle \tableofcontents {
\newpage
\section{Introduction}

{\bf "There are more things in heaven and earth, Horatio,\\
Than are dreamt of in your philosophy."}\\

{\it "Hamlet", W. Shakespear}\\

In our first Chapter we are going to present  some general notions
and ideas of modern theory of integrability trying to outline its
computability aspects. The reason of this approach is that though
this theory was wide and deeply developed in the last few decades,
its results are almost unusable for non-specialists in the area
due to its complexity as well as due to some specific jargon
unknown to mathematicians working in other areas. On the other
hand, a lot of known results are completely algorithmical and can
be used as a base for developing some symbolic programm package
dealing with the problems of integrability. Creation of such a
package will be of a great help not only at the stage of
formulating of some new hypothesis but also as a tool to get new
systematization and classification results, for instance, to get
complete lists of integrable equations with given properties as it
was done already for PDEs
with known symmetries \cite{sh1987}.\\

Some results presented here are quite simple and can be obtained
by any student acquainted with the basics of calculus (in these
cases direct derivation is given) while some ideas and results
demand deep knowing of a great deal of modern mathematics (in this
cases only formulations and references are given). Our main idea
is not to present here the simplest subjects of integrability
theory but to give its general description in simplest possible
form in order to give a reader a feeling what has been done
already and what could/should be done further in this area. Most
of the subjects mentioned here
 will be discussed in  details in the next Chapters.\\

We will use the word "integrability" as a generalization of the
notion "exactly solvable" for differential equations. Possible
definition of differential operator  will be discussed as well as
some definitions of integrability itself. Numerous examples
presented here are to show in particular that it is reasonable not
only to use different notions of integrability for different
differential equations but sometimes it proves to be very useful
to regard {\bf one equation} using  various definitions of
integrability, depending on what properties of the equation are
under the study. Two classical approaches to classification of
integrable equations - conservation laws and Lie symmetries  - are
also briefly presented and a few examples are given in order to
demonstrate deep difference between these two notions, specially
in case of PDEs. Two interesting semi-integrable systems are
introduced showing one more aspect of integrability theory - some
equations though not integrable in any strict sense, can be
treated  as "almost" integrable due to their intrinsic properties.

\section{Notion of differential operator}

There are many ways to define linear differential first-order
operator $D$, starting with Leibnitz formula for product
differentiating
$$
D(ab)=a^{'}b+ab^{'}.
$$
This definition leads to
$$
D\cdot a= a^{'}+aD, \quad D^2a=aD^2+2a^{'}D+a^{''},....,
$$
\be \label{dn} D^n \cdot a=\sum_{k} \left( \ba{c}n\\k \ea \right)
D^k(a)D^{n-k} \ee

where
$$
\left( \ba{c}n\\k \ea \right)=\frac{n(n-1)...(n-k+1)}{1\cdot 2
\cdots  k} \quad \mbox{with} \quad \left( \ba{c}n\\0 \ea \right)=1
$$
are binomial coefficients. Thus, {\bf one} linear differential
first-order operator and its powers are defined. Trying to define
a composition of two linear operators, we get already quite
cumbersome formula

\be \label{d1d2}D_1D_2a=D_1(D_2(a)+aD_2)=D_1(a)D_2+D_1D_2(a) +
D_2(a)D_1+aD_1D_2, \ee

 which can be regarded as a {\bf definition}
of a factorizable linear differential second-order operator.
Notice that though each of $D_1$ and $D_2$ satisfies Leibnitz
rule, their composition $D_1D_2$ {\bf does not!} An important
notion of commutator of two operators $[D_1D_2-D_2D_1]$ plays a
 role of special multiplication\footnote {see Ex.1}. In case of
{\bf non-factorizable} second-order operator an attempt to
generalize Leibnitz rule leads to a very complicated Bourbaki-like
constructions which we are not going to present here. All these
problems appear due to
coordinateness of this approach.\\

On the other hand, a linear differential operator being written in
coordinate form as
$$ L=\sum_j f_j\p_j $$
leads to
$$ L=\sum_{|\a|\leq m} f_{\a}\p^{\a}, \quad \p^{\a}= \p^{\a_1}\cdots\p^{\a_n},   \quad |\a|=\a_1+...+\a_n $$
which is  {\bf definition} of LPDO of order $m$ with $n$
independent variables. Composition of two operators $L$ and $M$ is
defined as
$$
L\circ M= \sum f_{\a}\p^{\a}\sum g_{\b}\p^{\b}=\sum h_{\g}\p^{\g}
$$\\
and coefficients $h_{\g}$ are to be found from formula
(\ref{dn}).\\

Now,  notion of linear differential {\bf equation} (LDE) can be
introduced in terms of the kernel of differential operator, i.e.
$$
\ker (L):= \{\varphi| \sum_{|\a|\leq m} f_{\a}\p^{\a}\varphi =0\}.
$$

Let us regard as illustrating example second-order LODO with one
independent variable $x$:
$$ L= f_0+f_1\p+f_2\p^2, \quad \p:=\frac{d}{dx}. $$

Notice that if two functions $\varphi_1$ and $\varphi_2$ belong to
its kernel $\ker (L)$, then
$$
c_1 \varphi_1 + c_2 \varphi_2 \in \ker (L),
$$
i.e. $\ker (L)$ is a linear vector space over constants´ field
(normally, it is $\Real$ or $\Complex$).\\

Main theorem about ODEs states the existence and uniqueness of the
Cauchy problem for any ODE, i.e. one-to-one correspondence between
elements of the kernel and initial data. In our case Cauchy data
$$
\varphi |_{x=x_0}=\varphi^0, \quad \p \varphi |_{x=x_0}=\varphi^1
$$
form a two-dimensional vector space and, correspondingly,
dimension of kernel is equal 2, $dim(\ker (L))=2$. Any two
functions $\varphi_1, \varphi_2 \in \ker (L)$ form its basis if
Wronskian $<\varphi_1, \varphi_2>$ is non-vanishing:
$$
<\varphi_1, \varphi_2>:= \left| \ba{cc}\varphi_1 \
\varphi_2\\\varphi_1^{'} \ \varphi_2^{'}\ea \right| \ne 0
$$
while an arbitrary function $\psi \in \ker (L)$ has to satisfy
following condition:
$$
<\varphi_1, \varphi_2, \psi>:= \left| \ba{ccc}\varphi_1 \
\varphi_2 \ \psi \\ \varphi_1^{'} \ \varphi_2^{'}\ \psi^{'}\\
 \varphi_1^{''} \ \varphi_2^{''}\ \psi^{''}  \ea
\right| = 0.
$$
Now we can construct immediately differential operator with a
given kernel as
$$
L(\psi)=<\varphi_1,\varphi_2, \psi>.
$$
For instance, if we are looking for an LODE with solutions $\sin
{x}$ and $\sqrt{x}$, then
 corresponding LODE has form
$$
\psi^{''}(1-\frac 12 \tan {x} ) + \tan {x} \psi^{'} - \frac{1}{2x}
\psi - \frac 34 \frac{1}{x^2} \psi=0.
$$

Coming back to LODO of order $m$, we re-write formula for the
kernel in the form

\be \label{kern} L(\psi)=\frac{<\varphi_1,..., \varphi_m,
\psi>}{<\varphi_1,...,\varphi_m>} \ee which provides that
high-order coefficient $f_m=1$. It is done just for our
convenience and we will use this form further.\\

Now, some constructive definition of linear differential operator
was given and importance of its kernel was demonstrated.
Corresponding differential equation was defined in terms of this
kernel and construction of operator with a given kernel was
described. All this is {\bf not possible} for nonlinear operator
because in this case manifold of solutions has much more
complicated structure then just linear vector space - simply
speaking, the reason of it is that in this case linear combination
of solutions is not a solution anymore.  Due to this reason some
other notions are to be used to study properties of nonlinear
operators - symmetries, conservation laws and, of course, as the
very first step - change of variables transforming a nonlinear
operator into a linear one. We will discuss all this in the next
sections.

\section{Notion of integrability}
\begin{itemize}

\item{}{\bf 3.1. Solution in elementary functions:}
$$ \boxed{y''+y=0.}$$
 General solution of this equation belongs to the class of
trigonometrical  functions, $y= a \sin( x + b)$, with arbitrary
const $a,b$. In order to find this solution one has to notice that
this equation is LODE with constant coefficients which possess
fundamental system of solutions, all of the form $e^{\lambda x}$
where $\lambda$ is a root of characteristic polynomial.

\item{}{\bf 3.2. Solution {\it modulo} class of functions:}
\begin{itemize}
\item[]{\bf 3.2.1.}
$$\boxed{y^{''}=f(y).}
$$

In order to integrate this equation let us notice that
$y^{''}y'=f(y)y'$ which leads to
$$
\frac12 y'^2=\int f(y) dy + const = F(y)
$$

and finally

$$
   {dx}=\frac {dy}{\sqrt{2F(y)}},
$$

which describes differential equation with {\bf separable
variables}. In case when $F(y)$ is a polynomial of third or fourth
degree, this is {\bf definition of elliptic integral} and
therefore the initial nonlinear ODE is integrable in elliptic
functions. Particular case when polynomial $F(y)$  has multiple
roots might leads to a particular solution in elementary
functions. Let us regard, for instance, equation $y''= 2y^3$ and
put $const=0$, then $y'^2=y^4$ and $y=1/x$, i.e. we have ONE
solution in the class of {\it rational functions}. General
solution is written out in terms of {\it elliptic functions}.
Conclusion: equation is {\bf integrable} in the class of elliptic
functions and {\bf not integrable} in the class of
rational functions.\\

Notice that as a first step in finding of solution, the order $n$
of initial ODE was diminished to $n-1$, in our case $n=2$ . Of
course, this is {\bf not possible} for any arbitrary differential
equation. This new ODE is called {\bf first integral} or {\bf
conservation law} due its physical meaning in applications, for
instance, our example can be reformulated as Newton second law of
mechanics and its first integral corresponds to energy
conservation law.\\

\item[]{\bf 3.2.2.}
$$\boxed{y''=y^2+x.}$$

This equation defines first {\bf Painleve transcendent}. About
this equation  it was proven that it has no solutions in classes
of elementary or special functions. On the other hand, it is also
proven that {\bf Painleve transcendent} is a meromorphic function
with known special qualitative  properties (\cite{p3}).\\

This example demonstrates us the intrinsical difficulties when
defining the notion of integrability. Scientific community has no
general opinion about integrability of {\bf Painleve
transcendent}. Those standing on the classical positions think
about it as about non-integrable equation. Those who are working
on different applicative problems of theoretical physics involving
the use of {\bf Painleve transcendent} look at it as at some new
special function .
\end{itemize}

\item{}{\bf 3.3. Solution} {\it modulo} {\bf inexplicit function:}\\
 $$\boxed{u_t=2uu_x.}$$

This equation describes so called {\bf shock waves} and its
solutions are expressed in terms of inexplicit function.  Indeed,
let us rewrite this equation in the new independent variables
$\tilde{t}=t$, $\tilde{x}=u$ and dependent one $\tilde{u}=x$, i.e.
now $x=\theta (t,u)$ is a function on $t,u$. Then

$$d\tilde{t}=dt, \quad dx= \theta_t dt + \theta_u
du, \quad u_t |_{dx=0}=-\frac{\theta_t}{\theta_u}, \quad u_x
|_{dt=0}=\frac{1}{\theta_u}$$ and
$$
-\frac{\theta_t}{\theta_u}=2 u \frac{1}{\theta_u} \Rightarrow
-\theta_t= 2u \Rightarrow -\theta=2ut-\varphi(u)
$$
and finally  $ x+2tu= \varphi(u) $ where $\varphi(u)$ is {\bf
arbitrary function} on $u$. Now, we have finite answer but no
explicit form of dependence $u=u(x,t)$. Has the general solution
been found? The answer is that given some initial conditions, i.e.
$t=0$, we may define solution as $u=\varphi^{-1}(x)$ where is
$\varphi^{-1}$ denotes {\bf inverse function} for $\varphi$.

\item{}{\bf 3.4. Solution {\it modulo} change of variables
(C-Integrability):}

$$ \boxed{\psi_{xy} +\alpha \psi_x + \beta \psi_y + \psi_x  \psi_y =0 .}$$

This equation is called  {\bf Thomas equation} and it
 could be made linear with a change of variables. Indeed, let
$\psi= log \theta$ for some positively defined function $\theta$:
$$
\psi_x  = (log \theta)_x = \frac{\theta _x}{\theta},\quad \psi_y =
(log \theta)_y = \frac{\theta _y}{\theta}, $$$$ \psi_{xy}  = (
\frac{\theta_x}{\theta})_y = \frac {\theta_{xy} \theta  - \theta_y
\theta_x}{ \theta ^ 2}, \quad \psi_x  \psi _y = \frac {\theta_x
\theta_y}{\theta ^ 2}$$ and substituting this into Thomas equation
we get finally linear PDE

$$\boxed{\theta_{xy} + \alpha \theta_x + \beta \theta_y = 0.}$$

Suppose for simplicity that $\b=0$ and  make once more change of
variables: $\theta=\phi e^{k_1y}$, then
$$
\theta_x= \phi_x e^{k_1y}, \quad \theta_{xy}= \phi_{xy}
e^{k_1y}+k_1\phi_{x} e^{k_1y} \quad  \mbox{and} \quad  \phi_{xy} +
(k_1+\a) \phi_x  = 0,
$$
and finally
$$
\p_x (\phi_y + (k_1+\a) \phi)  = 0,
$$
which yields to
$$
\phi_y -k_2 \phi=f(y) \quad  \mbox{with} \quad k_2=-(k_1+\a)
$$
and arbitrary function $f(y)$. Now
 general solution can be obtained by the method
of variation of a constant. As a first step let us solve
homogeneous part of this equation, i.e. $ \phi_{y} - k_2\phi = 0$
and $\phi(x,y)= g(x)e^{k_2y}$ with arbitrary $g(x)$. As a second
step, suppose that $g(x)$ is function on $x,y$, i.e. $g(x,y)$,
then initial equation takes form
$$ (g(x,y)e^{k_2y})_{y} -
k_2g(x,y)e^{k_2y} = f(y),$$
$$g(x,y)_{y}e^{k_2y}+k_2g(x,y)e^{k_2y}- k_2g(x,y)e^{k_2y}=f(y),
$$
$$
g(x,y)_{y}e^{k_2y}=f(y), \quad g(x,y)= \int f(y)e^{-k_2y}dy +
h(x),
$$
and finally the general solution of Thomas equation with $\a=0$
can be written out as

$$
\phi(x,y)= g(x,y)e^{k_2y}=e^{k_2y}(\int f(y)e^{-k_2y}dy + h(x))=
\hat{f}(y)+e^{k_2y}h(x)
$$
with two arbitrary functions $\hat{f}(y)$ and $h(x)$.

 \item{}{\bf 3.5. Solution {\it modulo} Fourier transform
(F-Integrability):}
 $$\boxed{u_t=2uu_x+ \varepsilon u_{xx}}$$

where $\varepsilon$ is a constant. This equation is called {\bf
Burgers equation} and it differs from Thomas equation studied
above where change of variables was local in a sense that solution
in each point  does not depend on the solution in some other
points of definition domain, i.e. local in $(x,y)$-space. This
equation can be transformed into $$u_t=2uu_x+ u_{xx}$$ by the
change of variables $\tilde{x}=\varepsilon x$, $u=\varepsilon
\tilde{u}$ and $\tilde{t}=\varepsilon^2 t$. We will use this form
of Burgers equation skipping tildes in order to simplify the
calculations below.
 To find solution of  Burgers equation one has to use Fourier transform
 which obviously
is nonlocal, i.e. here solution is local only in $k$-space. In
order to demonstrate it let us integrate it using notation $$\int
u dx=v,$$ then after integration
$$v_t=v_x^2+ v_{xx}=e^{-v}(e^{v})_{xx}$$ and change of variables
$w=e^{v}$, Burgers equation is reduced to the {\bf heat equation}
$$w_t=w_{xx}$$

 which is linear. Therefore, solutions of
Burgers equation could be obtained from solutions of heat equation
by the change of variables $$u=v_x=\frac{w_x}{w}$$ as
$$
w(x,0)=\int_{-\infty}^ {\infty}\exp(ikx)\hat{w}(k)dk \Rightarrow
u(x,t)=\frac{\int \exp(ikx-k^2t)\hat{w}(k)ikdk}{\int
\exp(ikx-k^2t)\hat{w}(k)dk}$$

where $w(x,0)$ is initial data  and $\hat{w}(k)$  is a function
called {\bf its Fourier transform} and it can be computed as

$$
\hat{w}(k)=\frac{1}{2\pi}\int_{-\infty}^
{\infty}\exp(-ikx)w(x,0)dx.
$$

In fact, it is well-known that {\bf any} linear PDE with constant
coefficients on an infinite line can be solved using as standard
basis $\{e^{ikx}| k \in \Real\}$  because they are eigenfunctions
of these operators. Thus, Thomas equation with $\a=0$ where the
general solution was found explicitly, is {\bf an exception} while
heat equation
demonstrates the general situation. \\

 \item{}{\bf 3.6. Solution {\it modulo} IST
(S-Integrability):}  $$\boxed{u_t=6uu_x+ u_{xxx}.}$$

This equation is called {\bf Korteveg-de Vries (KdV) equation} and
it is nonlinear PDE with nonconstant coefficients. In this case,
choosing set of functions $\{e^{ikx}\}$ as a basis is not helpful
anymore: Fourier transform does not simplify the initial equation
and only generates an infinite system of ODEs
on Fourier coefficients.\\

On the other hand, some new basis can be found which allows to
reduce KdV with rapidly decreasing initial data, $u\to 0,\,
x\to\pm\infty$, to the linear equation and to solve it. This new
basis can be constructed using solutions of {\bf linear
Schrödinger equation}

$$\boxed{\psi_{xx}+k^2\psi=u\psi}$$

where function $u$ is called {\bf potential} due to its origin in
quantum mechanics. Solutions of linear Schrödinger equation are
called {\bf Jost functions}, $\psi^\pm(t,x,k)$, with asymptotic
boundary conditions:

$$
\psi^\pm(t,x,k; u(x,t))e^{-\pm i(kx+k^3t)}\to 1,\quad x\to\pm
\infty.
$$

Jost function $ \phi (x,k)=\psi^+(t,x,k; u(x,t))e^{-i(kx+k^3t)}$
is defined by the integral equation
$$
\phi (x,k)=1+\int_x^\infty\frac{1-\exp[2k(x-x')]}{2k}
u(x')\phi(x',k)dx'
$$

with $t$ playing role of a parameter. Second Jost function is
defined analogously with integration over $[-\infty,x]$. Notice
that asymptotically for $x\to\pm \infty$ linear Schrödinger
equation
$$ \psi_{xx}+k^2\psi=u\psi \quad \mbox{is reduced to} \quad
\psi_{xx}+k^2\psi=0$$ as in case of Fourier basis $\{e^{ikx}| k
\in \Real\} $. It means that asymptotically their solutions do
coincide and, for instance, any solution of linear  Schrödinger
equation $$\psi\sim c_1e^{i(kx+k^3t)}+c_2 e^{-i(kx+k^3t)} \quad
\mbox{for} \quad x\to\infty.$$

It turns out that solutions of KdV can be regarded as potentials
of linear Schrödinger equation, i.e. following system of equations
$$
\begin{cases}
u_t=6uu_x+ u_{xxx},\\
\psi_{xx}+k^2\psi=u\psi
\end{cases}
$$

is consistent  and any solution of KdV can be written out as an
expansion of Jost functions which in a sense are playing role of
exponents $e^{ikx}$ in Fourier basis \cite{sh1979}.\\

On the other hand, there exists a major difference between these
two basis: Fourier basis is written out in explicit form {\it via}
one function while Jost basis is written out in inexplicit form
{\it via} two functions with different asymptotic properties on
the different ends of a line. The crucial fact here is that two
Jost functions are connected by simple algebraic equation:
$$
\psi^{-}(x,k,t)=a(k)\psi^{+}(x,-k,t)+b(k)e^{ik^3t}\psi^{+}(x,k,t)
$$

while it allows us to construct rational approximation of Jost
functions for given $a(k)$ and $b(k)$ and, correspondingly,
general  solution of KdV. The problem of reconstruction of
function $u$ according to $a(k), b(k)$ is called {\bf inverse
scattering problem} and this method, correspondingly, {\bf inverse
scattering transform (IST)}.\\

 \item{}{\bf 3.6. Solution {\it modulo} Dressing method
(D-Integrability):}  $$\boxed{iu_t=u_{xx} \pm |u|^2u.}$$ This
equation is called {\bf nonlinear Schrödinger equation} (NLS) and
it is very important in many physical applications, for instance,
in nonlinear optics. Dressing method is generalization of IST and
in this case role of auxiliary linear equation (it was linear
Schrödinger equation, second order ODE, in the previous case)
plays a system of two linear first order ODEs \cite{zak1971}:
$$
\begin{cases}
\psi_x^{(1)}=\lambda\psi^{(1)}+u \psi^{(2)}\\
\psi_x^{(2)}=-\lambda\psi^{(2)}+v \psi^{(1)}
\end{cases}
$$
where $v=\pm \bar{u}$. It turns out that system of equations
$$
\begin{cases}
iu_t=u_{xx} \pm |u|^2u\\
\psi_x^{(1)}=\lambda\psi^{(1)}+u \psi^{(2)}\\
\psi_x^{(2)}=-\lambda\psi^{(2)}+v \psi^{(1)}
\end{cases}
$$
is consistent and is equivalent to Riemann-Hilbert problem.
Solutions of this last system are called {\bf matrix Jost
functions} and any solution of NLS can be written out as an
expansion of matrix Jost functions which in a sense are playing
role of
exponents $e^{ikx}$ in Fourier basis \cite{sh1975}. Detailed presentation of Dressing
method will be given in some of our further Chapters.\\
\end{itemize}

\section{Approach to classification}

Our list of definitions is neither full nor exhaustive, moreover
one equation can be regarded as integrable due to a few different
definitions of integrability. For instance, equation for shock
waves from § {\bf 3.3}, $u_t=2uu_x$, is a particular form of
Burgers equation from § {\bf 3.5}, $u_t=2uu_x+ \varepsilon u_{xx}$
with $\varepsilon=0$, and it can be linearized as above, i.e. it
is not only integrable in terms of   inexplicit function but also
C-integrable and  F-integrable, with general solution
$$u(x,t)=\frac{\int \exp(ikx-k^2t)\hat{u}(\varepsilon k)ikdk}{\int
\exp(ikx-k^2t)\hat{u}(\varepsilon k)dk}.$$ What form of
integrability is chosen for some specific equation depends on what
properties of it we are interested in. For instance, the answer in
the form of inexplicit function shows immediately dependence of
solution form on initial conditions - graphically presentation of
inverse function $u=\varphi^{-1}(x)$ can be obtained as mirrored
image of $x=\varphi (u)$. To get the same information  from the
formula above is a very nontrivial task. On the other hand, the
general formula is the only known tool to study solutions with
singularities. This shows that definitions of integrability do not
suit to serve as a basis for classification of integrable systems
and some more intrinsic ways should be used to classify and solve
them. Below we present briefly two possible classification bases -
 conservation laws and Lie symmetries.

\subsection{Conservation laws}

Some strict and reasonable  definition of a {\bf conservation law}
(which is also called {\bf first integral} for ODEs) is not easy
to give, even in case of ODEs. As most general definitions one
might regard
\begin{equation}\label{clOd}
\frac{d}{dt}F(\vec{y})=0 \quad \mbox{for ODE} \quad
\frac{d}{dt}\vec{y}=f(\vec{y}), \quad  \vec{y}=(y_1,...,y_n),
\quad  f(\vec{y})=(f_1,...,f_n)
\end{equation}
 and
\begin{equation}\label{clPa}
\frac{d}{dt}\int G(u,u_x,u_y,u_{xx},u_{xy},u_{yy}...)dxdy...=0
\quad \mbox{for PDE} \quad \p_t u=
g(u,u_x,u_y,u_{xx},u_{xy},u_{yy},...).
\end{equation}
\\
Obviously, without putting some restrictions on function $F$ or
$G$ these definitions are too general and do not even point out
some specific class of differential equations. For instance, let
us take {\bf any} second order ODE, due to well-known theorem on
ODEs solutions we can write its general solution in a form
$$
F(t,y,a,b)=0
$$
where $a,b$ are two independent parameters (defined by initial
conditions). Theorem on inexplicit function gives immediately
$$
a=F_1(t,y,b) \quad \forall b\quad \mbox{and}\quad b=F_2(t,y,a)
\quad \forall a,
$$
i.e. any  second order ODE has 2 independent conservation laws and
obviously, by the same way  $n$ independent conservation laws can
be constructed for ODE of order $n$. For instance, in the simplest
case of second order ODE with constant coefficients, general
solution and its first derivative have form
$$
\begin{cases}
y=c_1e^{\lambda_1x}+c_2e^{\lambda_2x} \\
y^{'}=c_1\lambda_1e^{\lambda_1x}+c_2\lambda_2e^{\lambda_2x}
\end{cases}
$$
and multiplying $y$ by $\lambda_1$ and $\lambda_2$ we get
equations on $c_1$ and $c_2$ correspondingly:
$$
\begin{cases}
\lambda_2 y- y^{'}=c_1(\lambda_2-\lambda_1)e^{\lambda_1x} \\
\lambda_1 y- y^{'}=c_2(\lambda_1-\lambda_2)e^{\lambda_2x}
\end{cases} \Rightarrow \quad
\begin{cases}
\lambda_2x + \hat{c_2}=log(y^{'}-\lambda_1 y )\\
\lambda_1x + \hat{c_1}=log(y^{'}-\lambda_2 y )
\end{cases}
$$
and two conservation laws are written out explicitly:
$$
\hat{c_2}=log(y^{'}-\lambda_1 y )-\lambda_2x, \quad
\hat{c_1}=log(y^{'}-\lambda_2 y )-\lambda_1x.
$$
To find these conservation laws without knowing of solution is
more complicated task then to solve equation itself and therefore
they give no additional information about equation. This
 is the reason why often only polynomial or rational
conservation laws are regarded - they are easier to find and
mostly they describe qualitative  properties
of the equation which are very important for applications (conservation of energy, momentum, etc.) \\

On the other hand, conservation laws, when known, are used for
construction of ODEs solutions. Indeed, let us rewrite
Eq.(\ref{clOd}) as
$$
\frac{d}{dt}F(\vec{y})=(f_1\p_1 + f_2\p_2 + ... + f_n
\p_n)F=\mathcal{L}(F)=0,
$$
i.e. as an equation in partial derivatives  $\mathcal{L}(F)=0$.
Such an equation has
 $(n-1)$ {\bf independent}
particular solutions $(\varphi_1, \varphi_2,...,\varphi_{n-1})$ if
its  Jacobian matrix has  maximal rank
 \be rank  \frac{\p(\varphi_1, \varphi_2,...,\varphi_{n-1})}{\p (y_1, y_2,...,y_{n-1})} =n-1,\ee
 with notation
$$
\frac{\p(\varphi_1, \varphi_2,...,\varphi_{n-1})}{\p (y_1,
y_2,...,y_{n-1})}= \left( \ba {cccc} \frac{\p \varphi_1}{\p y_1} \
\frac{\p
\varphi_1}{\p y_2} \ \cdots \ \frac{\p \varphi_1}{\p y_n}\\
\cdots \ \cdots \ \cdots \ \cdots \\
\frac{\p \varphi_n}{\p y_1} \ \frac{\p \varphi_n}{\p y_2} \ \cdots
\ \frac{\p \varphi_n}{\p y_n} \ea \right).
$$

Now we can write out the general solution as $F(\varphi_1,
\varphi_2,...,\varphi_{n-1})$ with arbitrary function $F$ and
 initial ODE can be reduced to
$$
\frac{dz}{dt}=f(z), \quad \frac{dz}{f(z)}=dt
$$
and solved explicitly in quadratures (see § {\bf 3.2}).

\subsection{Symmetry properties}
In order to give  definition of
 canonical form for $n$-order ODE

\be \label{canon}
 y^{(n)}=F(x,y,y^{'},...,y^{(n-1)})
\ee
 let us first introduce vector

$$ \vec{y}=\begin{bmatrix}x\cr y\cr y^{'}\cr...\cr
y^{(n-1)}\end{bmatrix}
$$
which is called {\bf vector of dynamical variables} with all its
coordinates regarded as independent, and its first derivative
$$
\frac{d\vec{y}}{dt}=\begin{bmatrix}1\cr y^{'}\cr y^{''}\cr...\cr
y^{(n)}\end{bmatrix} =\begin{bmatrix}1\cr y^{'}\cr y^{''}\cr...\cr
F\end{bmatrix}
$$
with respect to some new independent variable $t$ such that
$dt=dx$, then the equation
\begin{equation}\label{dynamic}
\frac{d\vec{y}}{dt}=\begin{bmatrix}1\cr y^{'}\cr y^{''}\cr...\cr F
\end{bmatrix}
\end{equation}
is called {\bf canonical form} of an ODE. This canonical form is
also called {\bf dynamical system}, important fact is that
dimension of dynamical system is $ (n+1)$ for $n$-order ODE.\\

\paragraph{Definition 4.1.} Dynamical system
\begin{equation}\label{symm}
\frac{d\vec{y}}{d \tau}=g(\vec{y})
\end{equation}
is called {\bf a symmetry} of  another dynamical system
\begin{equation*}
\frac{d\vec{y}}{d t}=f(\vec{y}),
\end{equation*}
if
\begin{equation}\label{cond}
\frac{d}{d \tau}(\frac{d\vec{y}}{d t})=\frac{d}{d
t}(\frac{d\vec{y}}{d \tau})\quad  \Leftrightarrow \quad \frac{d}{d
\tau}(f(\vec{y})=\frac{d}{d t}(g(\vec{y}))
\end{equation}
holds. Symmetry $g$ of dynamical system $f$ is called {\bf trivial} if $g=\const \cdot f$.\\

 Obviously, Eq.(\ref{cond}) gives necessary condition of
 compatibility of this two dynamical systems. It can be proven
 that this condition is also sufficient. Therefore, construction
 of each Lie symmetry with group parameter
$\tau$ is equivalent to a construction
 of some  ODE which have $\tau$
as independent variable and is consistent with a given ODE. System
of ODEs obtained this way, when being written in canonical form
 is called {\bf dynamical system
connected to the element of Lie symmetry algebra}. \\

 We will discuss it in all
details later (also see \cite{Olver}), now just pointing out the
fact that Lie algebra can be generated not only by Lie
transformation group (normally used for finding of solutions) but
also by the set of dynamical systems (\ref{dynamic}) (normally
used for classification purposes).\\

Two following elementary theorems show interesting
interconnections  between conservation laws and symmetries in case
of ODEs.\\

\paragraph{Theorem 4.1}{\it Let dynamical system
\begin{equation*}
\frac{d\vec{y}}{d \tau}=g(\vec{y})
\end{equation*}
is a symmetry of  another dynamical system
\begin{equation*}
\frac{d\vec{y}}{d t}=f(\vec{y})
\end{equation*}}
and $$ \frac{d}{dt}F(\vec{y})=(f_1\p_1 + f_2\p_2 + ... + f_n
\p_n)F=\mathcal{L}(F)=0
$$
is a conservation law. Then $Fg$ is symmetry as well.\\

$\blacktriangleright$ Indeed, let us introduce
$$
\mathcal{M}=(g_1\p_1 + g_2\p_2 + ... + g_n \p_n),
$$
then consistency condition Eq.(\ref{cond}) is written out as

\be \label{cons} \mathcal{L} \circ \mathcal{M}=\mathcal{M} \circ
\mathcal{L} \ee

 and after substituting $F\mathcal{M}$ instead of
$\mathcal{M}$ we get on the left hand of (\ref{cons})
$$
\mathcal{L}(F\mathcal{M})=\mathcal{L}(F)\mathcal{M}+F\mathcal{L}\mathcal{M}
=F\mathcal{L}\mathcal{M}=F\mathcal{M}\mathcal{L},
$$
and on the right hand of
$$
(F\mathcal{M})\mathcal{L}=F\mathcal{M}\mathcal{L}.
$$
\qed

\paragraph{Corollary 4.2}{\it  Let $n$-th order ODE  of the form (\ref{canon})} has a symmetry
$$
\frac{d\vec{g}}{dt}=\begin{bmatrix}g_0 \cr g_1\cr g_2\cr...\cr
g_n\end{bmatrix},
$$
then $g_0$ is a conservation law and consequently without loss of
generality we may put $g_0=1$, if $g_0 \ne 0$.

\paragraph{Theorem 4.3} {\it An ODE of arbitrary order $n$ having  of $(n-1)$
independent conservation laws (i.e. complete set), has no
nontrivial symmetries consistent with conservation laws}.\\

$\blacktriangleright$ Indeed, using all conservation laws we can
reduce original $n$-order ODE into first order ODE
$$
\frac{da}{dt}=f(a)
$$
and look for symmetries in the form
$$
 \frac{da}{d\tau}=g(a).
$$
Then
$$
\frac{d}{d\tau}(\frac{da}{dt})=f^{'}(a)g(a), \quad
\frac{d}{dt}(\frac{da}{d\tau})=g^{'}(a)f(a)
$$
and finally
$$
\frac{f^{'}}{f(a)}=\frac{g^{'}}{g(a)} \quad \Rightarrow \quad
ln(\frac{f}{g})= \const,
$$
i.e. functions $f(a)$ and $g(a)$ are proportional. \qed \\

 Simply speaking, Eq.(\ref{symm}) defines a one-parameter
transformation group with parameter $\tau$ which conserves the
form of original equation.  The very important achievement of Lie
was his first theorem giving constructive procedure for obtaining
such a group. It allowed him to classify integrable differential
equations and to solve them. The simplest example of such a
classification
  for  second order ODEs with two
symmetries
is following: each of them can be transformed into one of the four types\\

{\bf (I)} $y^{''}=h(y^{'})$  \quad {\bf (II)} $y^{''}=h(x)$, \quad
{\bf  (III)} $y^{''}=\frac{1}{x}h(y^{'})$, \quad
{\bf (IV)} $y^{''}=h(x)y^{'}$,\\

where $h$ denotes an arbitrary smooth function while explicit form
of corresponding transformations was also
written out by Lie.  \\

Symmetry approach can also be used for PDEs.  For instance,
first-order PDE for shock waves $u_t=2uu_x$ (§ {\bf 3.3}) has
following Lie symmetry algebra $u_{\tau}=\varphi(u) u_x$ where any
smooth function $\varphi=\varphi(u)$ defines one-parameter Lie
symmetry group with corresponding choice of parameter $\tau$.
Indeed, direct check gives immediately
$$
(u_t)_{\tau}=(2\varphi u_x)u_x + 2u(\varphi u_{xx} +
\varphi^{'}u_x^2)
$$
and
$$
(u_{\tau})_t=\varphi^{'}u_x 2uu_x+\varphi(2u_x^2+2uu_{xx})
$$
while relation $(u_t)_{\tau}= (u_{\tau})_t$ yields to the final
answer (here notation $\varphi^{'}\equiv \varphi_u$ was used). In
order to construct dynamic system for this equation, let us
introduce dynamical independent variables as
$$
u, \quad u_1=u_x, \quad u_2=u_{xx}, \quad ...
$$
and  dynamical system as
$$
\frac{d}{dt}
\begin{bmatrix}u\cr u_1\cr u_2\cr ... \end{bmatrix}=
\begin{bmatrix}2uu_1\cr 2uu_2+2u_1^2 \cr 2uu_3+6u_1u_2\cr ... \end{bmatrix}.
$$
This system can be transformed into finite-dimensional system
using characteristics method \cite{sh1991}.\\

For PDE of order $n>1$  analogous dynamic system turns out to be
always infinite and only particular solutions are to be
constructed but no general solutions. Infinite-dimensional
dynamical systems of this sort are not an easy treat and
 also choice of dynamical variables presents sometimes a special
 problem to be solved, therefore even in such an exhaustive textbook as Olver's \cite{Olver}
 these systems are not even discussed. On the other hand,
 practically all known results on  classification of integrable
 nonlinear PDEs of two variables have been obtained
 using this approach (in this context the notion of F-Integrability is
 used as it was done to integrate Burgers equation, {\bf § 3.5.}) For instance, in \cite{zhi1979} for a PDE of
 the form
 $$
u_{xy}=f(x,y)
 $$
with arbitrary smooth function $f$ on the right hand it was proven
that this PDE is integrable and has symmetries {\bf iff} right
part has one of the following forms: $e^u$, $\sin{u}$ or
$c_1e^u+c_2e^{-2u}$ with arbitrary constants $c_1,c_2$. Another
interesting result was presented in \cite{sh1991}  where all PDEs
of the form
 $$
u_t=f(x, u , u_x, u_{xx})
 $$
have been classified. Namely, a PDE of this form is integrable and
has symmetries {\bf iff} if it can be linearized by some special
class of transformation. General form of transformation is written
out explicitly.

\subsection{Examples}
Few examples presented here demonstrate different
constellations of symmetries (SYM), conservation laws (CL) and solutions (SOL)
for a given equation(s).\\

{\bf 4.3.1: SYM +, CL +, SOL +}.\\

Let us regard a very simple equation $$y^{''}=1,$$ then its
dynamical system can be written out as
$$
 \frac{d\vec{y}}{dt} =\begin{bmatrix}1\cr
y^{'}\cr F\end{bmatrix}=
\begin{bmatrix} 1\cr y^{'}\cr 1\end{bmatrix} \quad \mbox{with}
\quad dt=dx
$$
 and its general solution is
$y=\frac{1}{2}x^2+c_1x+c_2$ with two constants of integration. In
order to construct conservation laws, we need to resolve formula
for solution with respect to the constants $c_1, c_2$:
$$
c_1=y^{'}-x, \quad c_2=y+\frac{1}{2}x^2-xy{'}.
$$
Now,  we look for solutions $F(\vec{y})$ of the equation
$$
\frac{d}{dt}F(\vec{y})=(\px  + y^{'} \py  + \p_{y^{'}}) F
=\mathcal{L}(F)=0
$$
with  $\mathcal{L}=\px  + y^{'} \py  + \p_{y^{'}}$. Direct check
shows that $F=y^{'}-x$ and $F=y+\frac{1}{2}x^2-xy{'}$ are
functionally independent solutions of this equation. Moreover,
general solution is an {\bf arbitrary function} of two variables
$$ F=F(y^{'}-x,y+\frac{1}{2}x^2-xy{'}),$$ for example,
$$F= (y^{'}-x)^2-2(y+\frac{1}{2}x^2-xy{'})= y^{'}-2y.$$
On the other hand, if there are no restriction on the function
$F$, the conservation laws may take some quite complicated form,
for instance,
$$
F=Arcsin(y^{'}-x)/(y+\frac{1}{2}x^2-xy{'})^{0.93}.$$

Now, that dynamical system, conservation laws and general solution
of the original equation have been constructed, let us look for
its  symmetry:
$$\quad g(\vec{y}): \quad
\frac{d\vec{y}}{d\tau}=g(\vec{y}),\quad g(\vec{y})=(g_1,g_2,g_3).
$$
Demand
of compatibility
$$
\frac{d}{d \tau}(f(\vec{y})=\frac{d}{d t}(g(\vec{y}))\quad
\mbox{is equivalent to} \quad \mathcal{L}(g_1)=\mathcal{L}(g_3)=0,
\quad \mathcal{L}(g_2)=g_3,
$$
and it can be proven that any linear combination of two vectors
$(1,0,0)$ and $(0,x,1)$ with (some) scalar coefficients provides
solution of compatibility problem. Thus, Lie symmetry group
corresponding to the vector $(1,0,0)$ is shift in $x$ while the
second vector $(0,x,1)$ corresponds to summing $y$ with particular
solutions of homogeneous equation.\\

{\bf 4.3.2: SYM --, CL +, SOL +}.\\

Let us regard as a system of ODEs
\begin{eqnarray}\label{top}
\begin{cases}
n_1\frac{da_1}{dt}=(n_2-n_3)a_2a_3\\
n_2\frac{da_2}{dt}=(n_3-n_1)a_1a_3\\
n_3\frac{da_3}{dt}=(n_1-n_2)a_1a_2
\end{cases}
\end{eqnarray}
with variables $a_i$ and constants $n_i$. This system is
well-known in physical applications - it describes three-wave
interactions of atmospheric planetary waves, dynamics of elastic
pendulum or swinging string, etc. Conservation laws for this
system are:
\begin{itemize}
\item{} Energy conservation law is obtained by multiplying the
$i$-th equation by $a_i$, $i=1,2,3$ and adding all three of them:
$$
n_1 a^{2}_{1}+n_2 a^{2}_{2}+n_3 a^{2}_{3}=\const.
$$

\item{} Enstrophy conservation law is obtained by multiplying the
$\it i$-th equation by $n_i a_i$, $i=1,2,3$ and adding all three
of them:
$$
n_1^2 a^{2}_{1}+n_2^2 a^{2}_{2}+n_3^2 a^{2}_{3}=\const.
$$
\end{itemize}

 Using these two conservation laws
one can easily obtain expressions for $a_2$ and $a_3$ in terms of
$a_1$. Substitution of these expressions into the first equation
of Sys.(\ref{top}) gives us
 differential equation  on $a_1$:
 $$
 (y^{'})^2=f(y) \quad \mbox{with} \quad y=a_1
 $$
 whose explicit solution is one of Jacobian
 elliptic functions while $a_2$ and $a_3$ are two other Jacobian elliptic
 functions \cite{kar2005}. In fact, Sys.(\ref{top}) is often regarded as one
 of possible definitions of Jacobian elliptic
 functions.\\

Notice that from {\bf Theorem 4.3} one can conclude immediately
that all symmetries consistent with conservation laws, are trivial.\\

{\bf 4.3.3: SYM +, CL $\pm$, SOL +}.\\

Let us regard heat equation
\begin{equation}\label{noCL}
\p_t u = \p_{xx} u,
\end{equation}
which generates solutions of Burgers equation (see {\bf § 3.5}).
 Direct check shows that Eq.(\ref{noCL}) is invariant due to
transformations $x=\tilde{x}\tau$ and $t=\tilde{t}\tau^2$ with any
constant $\tau$, i.e. dilation transformations constitute Lie
symmetry group for Eq.(\ref{noCL}).  Moreover, Eq.(\ref{noCL}) is
integrable and its solution
$$
u=\int \exp(ikx-k^2t)\hat{u}(k)dk \quad \mbox{with}\quad
\hat{u}(k)=\frac{1}{2\pi}\int_{-\infty}^
{\infty}\exp(-ikx)u(x,0)dx
$$\\
is obtained by Fourier transformation.\\

This example demonstrates also some very peculiar property - heat
equation (as well as Burgers equation) has {\bf only one
conservation law}:
$$
\frac{d}{dt}\int_{-\infty}^{\infty} u  dx =0.
$$
Nonexistence of any other conservation laws is proven, for
instance, in \cite{sh1991}.

\section{Semi-integrability}
\subsection{ Elements of integrability}

Dispersive evolution PDEs on compacts is our subject in this
subsection. In contrast to standard mathematical classification of
LPDO into hyperbolic, parabolic and elliptic operators there
exists some other classification - into dispersive and
non-dispersive operators - which is successfully used in
theoretical physics and {\bf is not complementary} to mathematical
one (for details see \cite{kar2005}). Let regard LPDE with
constant coefficients in a form
$$
P(\frac{\partial}{\partial t},\frac{\partial}{\partial x})=0
$$
where $t$ is time variable and $x$ is space variable, and suppose
that a {\bf linear wave}
$$
\psi (x)=\tilde{A}\exp{i(k x -\o t)}
$$
with constant amplitude $\tilde{A}$, wave number $ k$ and
frequency $\o$ is its solution. After substituting a linear wave
into initial LPDE we get $ P(-i\o, ik)=0$, which means that  $k$
and $\o$ are connected in some way:
there exist some function $f$ such, that $ f(\o, k)=0.$\\

This connection is called {\bf dispersion relation} and  solution
of the dispersion relation is called {\bf dispersion function},
$\o = \o(k)$. If  condition

$$
\frac {\p^2\o}{\p k^2} \neq 0
$$
holds, then initial LPDE is called {\bf evolution dispersive
equation} and it obviously is completely defined by dispersive
function.  All these definitions  can be easily reformulated for a
case of more space variables, namely $x_1, x_2,...,x_n$. In this
case linear wave takes form
$$
\psi (x)=\tilde{A}\exp{i(\vec{k}\vec{x} -\o t)}
$$
with {\bf wave vector} $\vec{k}=(k_1,....k_n)$ and space-like
variable $\vec{x}=(x_1,...,x_n)$. Then

$$
P(\frac{\partial}{\partial t},\frac{\partial}{\partial
x_1},...,\frac{\partial}{\partial x_n})=0,
$$

dispersion function can be computed from $ P(-i\omega,
ik_1,...,ik_n)=0 $ and the condition of non-zero second derivative
of the dispersion function takes a matrix form:

$$ \arrowvert \frac {\partial^2\omega}{\partial k_i \partial k_j}
\arrowvert \neq 0.
$$

Notice now that solutions of linear evolution dispersive PDE are
known {\bf by definition} and the reasonable question here is:
what can be found about solutions of {\bf nonlinear} PDE
$$
\mathcal{L}(\psi)=\mathcal{N}(\psi)
$$
with dispersive linear
part $\mathcal{L}(\psi)$ and some nonlinearity $\mathcal{N}(\psi)$?\\

Nonlinear PDEs of this form play major role in the theory of wave
turbulence and in general there is no final answer to this
question. Case of {\bf weak turbulence}, i.e. when nonlinearity
$\mathcal{N}(\psi)$ is regarded small in a sense that wave
amplitudes $A$ are small enough (smallness of an amplitude can be
strictly defined), is investigated in much more details. Two
qualitatively different cases have to be regarded:

\begin{itemize}
\item[] {\bf 1.} coordinates of wave vector are real numbers,
$\{\vec{k}=(k_1,....k_n) | k_i \in \Real\}$ (corresponds to
infinite space domain);

\item[] {\bf 2.}  coordinates of wave vector are integer numbers,
$\{\vec{k}=(k_1,....k_n) | k_i \in \Integer\}$ (corresponds to
compact space domain).
\end{itemize}

 In the
first case method of {\bf wave kinetic equation} has been
developed in 60-th (see, for instance, \cite{has1962}) and applied
for many different types evolution PDEs. Kinetic equation is
approximately equivalent to initial nonlinear PDE but has more
simple form allowing direct numerical computations of each wave
amplitudes in a given domain of wave spectrum. Wave kinetic
equation is an averaged equation imposed on a certain set of
correlation functions and it is in fact one limiting case of the
quantum Bose-Einstein equation while the Boltzman kinetic equation
is its other limit.  Some statistical assumptions have been used
in order to obtain kinetic equations and limit of its
applicability then is a very complicated problem which should be
solved separately for
each specific equation \cite{zak1999}.\\

 In the second case, {\bf exact solutions} in
terms of elliptic functions have been found \cite{kar1998}. More
precisely, it is proven that solving of initial nonlinear PDE {\bf
can be reduced} to solving a few small systems of ODEs of the form
\begin{eqnarray} \label{res}
\begin{cases}
\dot{A}_1= \mathfrak {C}_1 A_2 A_3, \\
\dot{A}_2= \mathfrak {C}_2 A_1A_3, \\
\dot{A}_3= \mathfrak {C}_3 A_1A_2,
\end{cases}
\end{eqnarray}
in case of quadratic nonlinearity,
\begin{eqnarray}
\begin{cases}
\dot{A}_1= \mathfrak {C}_1 A_2 A_3 A_4, \nonumber\\
\dot{A}_2= \mathfrak {C}_2 A_1A_3A_4, \nonumber\\
\dot{A}_3= \mathfrak {C}_3 A_1A_2A_4, \nonumber\\
\dot{A}_4= \mathfrak {C}_4 A_1A_2A_3, \nonumber
\end{cases}
\end{eqnarray}
in case of cubic nonlinearity and so on. Notice, that in contrast
to a linear wave with a constant amplitude
$\tilde{A}\neq\tilde{A}(t,\vec{x})$, waves in nonlinear PDE have
amplitudes $A_i$  {\bf depending on time}. It means that solutions
of initial nonlinear PDE have characteristic wave form as in
linear case but wave amplitudes are Jacobian elliptic functions on
time, $cn(T), dn(T)$ and $ sn(T)$. Notice that Sys.(\ref{res}) has
been studied in {\bf § 4.3.2} and its conservation laws were
found. Exact solutions of Sys.(\ref{res}) are
\begin{eqnarray}
\begin{cases}
A_1=b_1 cn(T/t_0 -\lambda), \nonumber \\
A_2=b_2 dn(T/t_0 -\lambda), \nonumber \\
A_3=b_3 sn(T/t_0 -\lambda), \nonumber
\end{cases}
\end{eqnarray}
and constants $b_i, t_0, \lambda$ are written out explicitly as
functions of initial values of waves´ amplitudes (see
\cite{kar2005} for details).\\

 These systems of ODEs providing exact solutions of  initial nonlinear
 evolution PDE are called {\bf elements of integrability}.
Some constructive procedure, {\bf Clipping method}, has been
developed
 \cite{kar1994} allowing to find all elements of integrability for a given
evolution PDE.

\subsection{Levels of integrability}

 Let us formulate classical three-body problem whose
 integrability attracted attention of many investigators beginning
 with Lagrange. Computing the mutual gravitational interaction of
 three masses is surprisingly difficult to solve and only two integrable cases were found.
For simplicity we regard three-body problem with all masses
 equal, then equations of motion take form

\begin{eqnarray}\label{3body}
\begin{cases}
\frac{d^2z_1}{d t^2}= z_{12}f_{12}+z_{13}f_{13}\\
\frac{d^2z_2}{d t^2}= z_{21}f_{12}+z_{23}f_{23}\\
\frac{d^2z_3}{d t^2}= z_{31}f_{13}+z_{32}f_{23}
\end{cases}
\end{eqnarray}

where $z_j$ is a complex number, $z_j= x_j+iy_j$, describing
coordinates of $j$-th mass on a plane, $f_{jk}$ is a given
function depending on the distance between $j$-th and $k$-th
masses (physically it is attraction force) while following
notations are used: $z_{jk}=z_j-z_k$ and $f_{jk}=f(|z_{jk}|^2)$.\\

This system admits following conservation laws:
\begin{itemize}
\item{} {\bf Velocity of center of masses is constant}\\
Summing up all three equations, we get
$$
\frac{d^2}{d t^2}(z_1+z_2+z_3)=0.
$$
This equality allows us to choose the origin of coordinate system
in such a way that  $$z_1+z_2+z_3=0$$ which simplifies all further
calculations significantly. That is the reason why till the end of
this section this coordinate system is used. Physically it means
that coordinate system is connected with masses center.

\item{} {\bf Conservation of energy}\\
Multiplying $j$-th equation by $\bar{z}^{'}_j$, summing up all
three equations and adding complex conjugate, we obtain on the
left
$$
\sum_{j=1}^{3}( \bar{z}^{'}_j z^{''}_j + z^{'}_j \bar{z}^{''}_j)=
\frac{d}{dt}\sum_{j=1}^{3} z^{'}_j \bar{z}^{'}_j.
$$
i.e. left hand describes derivative of kinetic energy.\\

On the right we have derivative of potential energy $\mathcal{U}$:
$$
\frac{d}{dt}\mathcal{U}=\frac{d}{dt}(F(|z_{12}|^2)+F(|z_{13}|^2)+F(|z_{23}|^2))
\quad \mbox{with notation} \quad F^{'}=f
$$
and finally energy conservation law takes form

$$
\sum_{j=1}^{3} z^{'}_j \bar{z}^{'}_j=
F(|z_{12}|^2)+F(|z_{13}|^2)+F(|z_{23}|^2) + \const
$$
\item{}{\bf Conservation of angular momentum}

By differentiating of angular momentum $$\im \sum_{j=1}^{3}
z^{'}_j \bar{z}_j $$ with respect to $t$, we get
$$
\im\Big( \sum
|z^{'}_j|^2+f_{12}z_{12}\bar{z}_{12}+f_{13}z_{13}\bar{z}_{13}+f_{23}z_{23}\bar{z}_{23}\Big)=$$$$
\im\Big( f_{12}|z_{12}|^2 +f_{13}|z_{13}|^2+f_{23}|z_{23}|^2
\Big)=0,
$$
\end{itemize}

while force $f$ is some real-valued function.\\

In general case there are no other conservation laws and the
problem is not integrable. On the other hand, one may look for
some  periodical solutions of Sys.(\ref{3body}) and try to deduce
the necessary conditions of periodicity. Importance of the
existence of periodical solutions was pointed out already by
Poincare and is sometimes even regarded as a {\bf definition} of
integrability - just as opposite case for a
chaos.\\

{\bf Theorem 5.1.} {\it If $f_{ij} > 0, \quad \forall i,j$
(so-called repulsive case), then Sys.(\ref{3body}) has no
periodical solutions.}\\

$\blacktriangleright$ Indeed, in case of periodical solution,
magnitude of inertia momentum
$$
\mathcal{Z}:=|z_{12}|^2+|z_{13}|^2+|z_{23}|^2
$$
should have minimums and maximums as sum of distances between
three masses. On the other hand,
$$
\frac{1}{2}\frac{d^2}{dt^2}(|z_{12}|^2+|z_{13}|^2+|z_{23}|^2)
=|\dot{z}_{12}|^2+|\dot{z}_{13}|^2+|\dot{z}_{23}|^2 +
f_{12}|z_{12}|^2+f_{13}|z_{13}|^2+f_{23}|z_{23}|^2 > 0,
$$
which contradicts to the fact that function $\mathcal{Z}$ has to
have different signs in the points of minimum and maximum. \qed \\

One interesting case - {\bf Poincare case} - though does not lead
to integrable reduction of Sys.(\ref{3body}), give quite
enlightening results and allows to regard this case as "almost"
integrable. In this case there exists one more conservation law -
conservation of inertia momentum
$$
|z_{12}|^2+|z_{13}|^2+|z_{23}|^2=\const
$$
and it is possible due to a special choice of function
 $f_{jk}=1/|z_{jk}|^4$ which allows us to reduce initial system to the
ODE of the form

\be \label{per}
\mathcal{B}^{''}=a(\mathcal{B}^{'})^3+b(\mathcal{B}^{'})^2+c\mathcal{B}^{'}+d.
\ee

in new polar coordinates. This equation describes a geometrical
place of points  on the plane, i.e. some plane curve
$\mathcal{B}$, providing solutions of initial system. The curve
$\mathcal{B}$ is of figure-eight form and can not be described by
any known algebraic curve. On the other hand, it can be
approximated with desirable accuracy, for instance, by lemniscate
$$
x^4+\a x^2y^2+\b y^4=x^2-y^2.$$ Very comprehensive collection of
results and graphics one can find in \cite{fu}\\

There exists hypothesis that  {\bf the only periodical solution}
of Eq.(\ref{per}) is this eight-form curve (not proven). Existence
theorem for non-equidistant  periodical solutions is proven for a
wide class of functions $f$ (in variational
setting). \\

The simplest possible case of periodical solution can be obtained
if one of  $z_i$ is equal to zero\footnote{see Ex.6} (obviously,
the problem is reduced to two-body problem). Two more complicated
classical integrable cases with periodical solutions for
particular choice of $z$-s are known:

\begin{itemize}
\item [--] {\bf Lagrange case}:
 $|z_{12}|=|z_{13}|=|z_{23}|$.\\
 It means that  distances between three masses are
 equal as well as all corresponding  attraction forces, and the masses are
 moving along a circle. Lagrange case is also called equidistance case. In this case
Sys.(\ref{3body}) can be reduced to ODE $y^{''}=f(y)$ and solved
in quadratures.
 \item [--] {\bf Calogero case}: all $z_j$ are real and
 $f_{jk}=1/|z_{jk}|^4$ for $j,k=1,...n$\\
It means that all masses are moving along a line (in fact, along a
real axes) and this is generalization of Euler case of three-body
problem which after appropriate change of variables
 Sys.(\ref{3body})  takes form
$$
\begin{cases}
\frac{d^2x_1}{d t^2}= \frac{1}{(x_1-x_2)^3} +
\frac{1}{(x_1-x_3)^3}\\
\frac{d^2x_2}{d t^2}= -\frac{1}{(x_1-x_2)^3} +
\frac{1}{(x_2-x_3)^3}\\
\frac{d^2x_3}{d t^2}= -\frac{1}{(x_1-x_3)^3} -
\frac{1}{(x_2-x_3)^3}\\
\end{cases}
$$

Euler´s system was generalized in \cite{ca}, \cite{ca1} to
$$
\frac{d^2x_j}{d t^2}=\frac{\p \mathcal{U}}{\p x_j} \quad
\mbox{with} \quad \mathcal{U}=\sum_{i<j}\frac{1}{(x_i-x_j)^2}
\quad \mbox{and} \quad j=1,...,n.
$$
All $n$ independent conservation laws were found and it was proven
that the system is integrable.
\end{itemize}

\section{Summary}

In our first Chapter we introduced a notion of differential
operator and gave few different definitions of integrable
differential equation. It was shown that some of them can be
equivalent for a given equation and it is reasonable to choose an
appropriate one depending on what properties of the equation are
under the study. Some interesting and physically important
examples of "almost" integrable systems were described. Very
intrinsic question on interconnections of conservation laws
and symmetries was also discussed. \\

Simplest possible example of nonlinear equation -  famous Riccati
equation which is ODE of first order with quadratic nonlinearity.
We will use this equation in our next Chapter, {\bf Chapter 2:
Riccati equation}, in order to demonstrate many properties of
differential equations described above. Riccati equation will also
be very useful for introduction of some new notions like
singularities of solutions, integrability tests, etc.

\section{Exercises for Chapter 1}
\paragraph{1.} Using (\ref{d1d2}) prove that $D_3=D_2D_1-D_1D_2$
satisfies Leibnitz rule.
\paragraph{2.} Prove that
operator $$\mathcal{L}=\sum a_k x^k \p_x^k$$ can be transformed
into an operator with {\bf constant} coefficients by the following
change of variables: $x=e^t$. (Euler)
\paragraph{3.}
Transform equation $\quad u_t=\varphi (u) u_x \quad$ into $ \quad
v_t=vv_x \quad$ by appropriate change of variables.
\paragraph{4.}
Prove that LODE with constant coefficients has
conservation laws of the form
$$
\frac{y^{'}-\lambda_1 y}{\lambda_2}e^{-\lambda_3 x}=\const.
$$
\paragraph{5.} For potential energy $\mathcal{U}$ from {\bf § 5.2}
prove that (Lagrange-Jacobi identity)
$$
x_1\frac{\p \mathcal{U}}{\p x_1} + y_1\frac{\p \mathcal{U}}{\p
y_1}+ x_2\frac{\p \mathcal{U}}{\p x_2}+y_2\frac{\p \mathcal{U}}{\p
y_2} + x_3\frac{\p \mathcal{U}}{\p x_3}+y_3\frac{\p
\mathcal{U}}{\p y_3}=
2(f_{12}|z_{12}|^2+f_{13}|z_{13}|^2+f_{23}|z_{23}|^2).
$$
\paragraph{6.} For particular case $z_1=0$ and our choice of coordinate system solve
Sys.(\ref{3body}) explicitly.

\section*{Acknowledgements}
 Author$^{2}$
acknowledges support of the Austrian Science Foundation (FWF)
under projects SFB F013/F1304. Both authors are very grateful to
Roma TRE University and Lecce University for their hospitality
during preparing of this paper.

\end{document}